\documentclass[conference]{IEEEtran}
\IEEEoverridecommandlockouts
\usepackage{cite}
\usepackage{amsmath,amssymb,amsfonts}
\usepackage{algorithmic}
\usepackage{graphicx}
\usepackage[caption=false,font=footnotesize]{subfig}
\usepackage{textcomp}
\usepackage{xcolor}
\usepackage{booktabs}
\usepackage{url}
\def\BibTeX{{\rm B\kern-.05em{\sc i\kern-.025em b}\kern-.08em
    T\kern-.1667em\lower.7ex\hbox{E}\kern-.125emX}}

\graphicspath{{figures/}}

\begin{document}

\bstctlcite{IEEEexample:BSTcontrol}

\title{5G-EMERGE: An Experimental Evaluation of Media Content Delivery in a Distributed 5G Network}

\author{
    \IEEEauthorblockN{  D. Verde\IEEEauthorrefmark{4}\textsuperscript{*},
                        C. Roseti\IEEEauthorrefmark{2}\IEEEauthorrefmark{4},
                        F. Zampognaro\IEEEauthorrefmark{3}\IEEEauthorrefmark{4}, and
                        M. Luglio\IEEEauthorrefmark{2}\IEEEauthorrefmark{4}
                        \thanks{\textsuperscript{*}Corresponding Author. E-mail: verde@romars.tech.}}
    \IEEEauthorblockA{
        \IEEEauthorrefmark{2} \textit{Department of Electronics Engineering, University of Rome ``Tor Vergata''}, Rome, Italy \\
        \IEEEauthorrefmark{3} \textit{Link Campus University}, Rome, Italy \\
        \IEEEauthorrefmark{4} \textit{RomARS s.r.l.}, Rome, Italy }}

\maketitle

\begin{abstract}
Over-The-Top (OTT) video services deliver content over HTTP, in unicast, so that the traffic generated at the origin and across the network grows linearly with the number of users. Content Delivery Networks (CDNs) mitigate this issue by placing multiple content replicas or web caches closer to users, while 5G natively enables an even closer placement, that is in a local Data Network (DN), which can be conveniently deployed at the edge of the system. This paper presents and experimentally validates the media delivery strategies of 5G-EMERGE. This ecosystem combines satellite-enabled edges, a commercial CDN, caching, and multicast delivery within a 5G system. Using a distributed testbed, we measure the latency experienced by unicast and multicast clients and the traffic load on the backhaul link and at the edge server. Results show that a cache hit in the local part of the DN reduces the average Time-to-First-Byte (TTFB) to about 1.5~ms, compared to 30.4~ms for a CDN hit and 123.4~ms when fetching the content directly from the origin, highlighting the latency penalty introduced by misses and request coalescing in the CDN. Finally, we show that multicast delivery keeps the load on both the edge server and the backhaul link constant regardless of the number of users, thus enhancing the scalability of the content delivery service with respect to unicast.
\end{abstract}


\section{Introduction}
Video streaming is today dominated by Over-The-Top (OTT) platforms such as Netflix, Amazon Prime Video, and DAZN. Being OTT, these services are delivered over the public Internet in unicast via HTTP, typically through adaptive streaming protocols such as Dynamic Adaptive Streaming over HTTP (DASH) or HTTP Live Streaming (HLS). As a consequence, a dedicated flow is established for each user, even when thousands or millions of users are watching the same content at the same time (e.g., a live sports event). This behaviour may overload not only the origin server, but also access networks, backhauls, and, more generally, the whole Internet.
To solve this problem, OTT services typically rely on geographically distributed Content Delivery Networks (CDNs). A CDN replicates the content on a set of replica servers and balances the requests of $n$ users over $m$ replicas, thus improving the scalability of the service with the number of users. Moreover, replica servers are placed closer to the users, thus reducing the end-to-end latency between the clients and the server from which the content is fetched. Clearly, latency has a direct impact on the Quality of Experience (QoE). As a reference, a large-scale study on Akamai traces showed that viewers start abandoning a video when the startup delay exceeds 2~s, and that every additional second increases the abandonment rate by 5.8\%~\cite{krishnan2013video}.
In this direction, 5G allows the deployment of a local Data Network (DN) at the network edge, in proximity to the end users~\cite{ts23501}. A DN can host an Application Server (AS) for specific local services (including replica servers and local web caches), and users could fetch AS content through their User Equipment (UE). The 5G Mobile Network Operator (MNO) could also restrict these services, making them accessible only to a subset of selected users (e.g., on a pay-per-view basis). In addition, 5G ensures that specific traffic flows receive dedicated Quality of Service (i.e., QoS enforcement), helping to reduce network congestion and avoid the traffic fluctuations that are typical of best-effort networks such as the Internet. Finally, 5G is access agnostic: the 5G Core Network (CN) is accessible not only through the standard Radio Access Network (RAN), but also through non-3GPP access networks, such as Wi-Fi~\cite{ts24502}.
In this paper, we present 5G-EMERGE\footnote{https://www.5g-emerge.com/technology}, an integrated ecosystem for media content delivery. 5G-EMERGE combines the benefits of CDNs, web caching, satellite communication, edge computing, and multicast, leveraging 5G as a pivotal technology. The main contributions of this work are the following:
\begin{itemize}
    \item Document the 5G-EMERGE content delivery strategies, mainly in compliance with (i)~the 5G Edge Computing framework~\cite{ts23548} and (ii) the 5G Multicast/Broadcast Services (MBS)~\cite{ts23247};
    \item Experimentally evaluate, on a distributed testbed with 5G non-3GPP access, the latency experienced by users when content is served by the origin server, a commercial CDN, and a web cache in the DN;
    \item Show the reduction in terms of traffic load (i) on the backhaul link and (ii) at the edge server achieved with multicast delivery from a satellite-fed edge server in the 5G DN, with respect to unicast.
\end{itemize}
Thus, the remainder of the paper is organized as follows. 
Section~\ref{sec:related} reviews related work in the literature. Section~\ref{sec:5g-emerge} summarizes the 5G-EMERGE architecture and its content delivery strategies. 
Section~\ref{sec:5g} describes the support for edge computing, non-3GPP access, and multicast/broadcast services in 5G. Section~\ref{sec:testbed} presents the evaluation testbed, while Section~\ref{sec:results} discusses the results achieved. Finally, Section~\ref{sec:conclusion} concludes the study and provides possible insights for future work.

\section{Related Work}\label{sec:related}
Multi-access Edge Computing (MEC) brings computing and storage resources close to the users, reducing access latency~\cite{carota2024easdf,taleb2017mec}. Researchers have extensively investigated how caching popular content at the edge of 5G networks helps to reduce latency and backhaul load~\cite{wang2014cache,bastug2014edge}, and have also analyzed the interplay between CDN caching and video streaming performance~\cite{ghabashneh2020cdn}. 
Satellite-assisted CDNs are discussed in~\cite{luglio2019satcdn,evans5gcache}, while possible satellite integration models in 5G are introduced in \cite{luglio2023modes,verde2022untrustedsatellite}. The use of multicast for edge caching in 5G-satellite integrated networks is studied in~\cite{wang2020codedmulticast}. 
In~\cite{carota2024easdf}, the authors show that resolving the Domain Name System (DNS) of MEC applications to closer servers during UE mobility reduces the 5G access latency with an overhead below 4~ms. In our previous work ~\cite{zampognaro2024les}, we observed that deploying Local Edge Services (LES) reduces the latency perceived by users connected over the 5G non-3GPP access. On the other hand, the delivery of multicast and broadcast services over 5G non-3GPP access is addressed in ~\cite{luglio2025mbsn3gpp,survey-5g-mbs}.
The performance of 5G multicast and broadcast services is evaluated through simulations in~\cite{ibanez2021mbs}, while some on-field test trials are reported in~\cite{pandolfi2025broadcast,china-field-test}. Finally, the delivery of multicast services in a distributed 5G network within 5G-EMERGE is preliminarily assessed through synthetic traffic in~\cite{verde2026multicast}. In this work, we focus instead on real media streaming applications and services.

\section{The 5G-EMERGE Ecosystem}\label{sec:5g-emerge}
\subsection{Overview}\label{sec:arch}
5G-EMERGE\footnote{https://resilience.esa.int/archives/projects/5gemerge} is a research project funded by the European Space Agency (ESA), which aims to design and validate an open and multi-tenant ecosystem that seamlessly distributes content from the origin server to many edge nodes, leveraging 5G as convergence technology. A detailed description of the reference architecture is provided in~\cite{verde2026multicast}. In 5G-EMERGE, edge nodes are classified according to their position with respect to the data origin:
\begin{itemize}
    \item \textbf{Near Edge}: a first-hop edge located close to the content origin. When the content provider publishes data to the origin server, the uplink edge makes it available to the other edge nodes;
    \item \textbf{Far Edge}: the edge closest to the final users, which can host replica servers and local caching services, serving content transparently to end users.
\end{itemize}
Near and far edges may be interconnected through a terrestrial or satellite backhaul network. For efficiency reasons, content is delivered mainly over satellite through a one-way multicast transmission based on Digital Video Broadcasting (DVB). In particular, 5G-EMERGE adopts DVB Multicast Adaptive Bit Rate (DVB-MABR)~\cite{dvbmabr}; in this way, the same content reaches many far edges at once, ensuring scalability of the ecosystem in terms of the number of edge nodes.
Within the far edge, users can connect via a 5G New Generation Radio Access Network (NG-RAN), or through Wi-Fi as a (untrusted or trusted) non-3GPP access network~\cite{ts23501,ts24502}. Since Wi-Fi operates in unlicensed bands, it can cost-effectively extend 5G coverage in confined or hard-to-reach environments, while preserving main 5G features such as user authentication and QoS enforcement.

\subsection{Content Delivery Strategies}\label{sec:strategies}
In 5G-EMERGE, the following content delivery optimization strategies are considered.

\textbf{CDN.} A CDN is an overlay network of replica servers, or Points of Presence (PoPs), deployed in many geographical locations~\cite{nygren2010akamai}. User requests are redirected, typically via DNS, to the most suitable PoP based on proximity and load. If the requested object is available (\textit{CDN hit}) it is served directly by the PoP; otherwise (\textit{CDN miss}) the PoP fetches it from the origin, stores and forwards it to the user. Therefore, at the same time a CDN (i)~offloads the origin, (ii)~balances the load among replicas, and (iii)~reduces the latency towards users, transparently.
    
\textbf{Web Cache.} A web cache (e.g., an HTTP reverse proxy) is a component that stores HTTP responses and serves repeated requests directly from its local storage, reducing (i)~end-to-end latency and (ii)~backend load. In the 5G System (5GS), a cache can be deployed directly in the local part of the DN (i.e., one hop away from the UPF). Thus, compared to a CDN PoP (which is typically located on the Internet), a local DN cache is located at the closest possible point to the users within the 5GS. Hence, a \textit{cache hit} minimizes user-perceived latency and prevents traffic from crossing the backhaul link between the edge and the rest of the network, which is particularly valuable when the backhaul is bandwidth-constrained or metered~\cite{wang2014cache,bastug2014edge}.

\section{The 5G System (5GS)}\label{sec:5g}
\subsection{Multi-Access Edge Computing in 5G}\label{sec:edge}
\begin{figure}[t]
    \centering
    \includegraphics[width=\columnwidth]{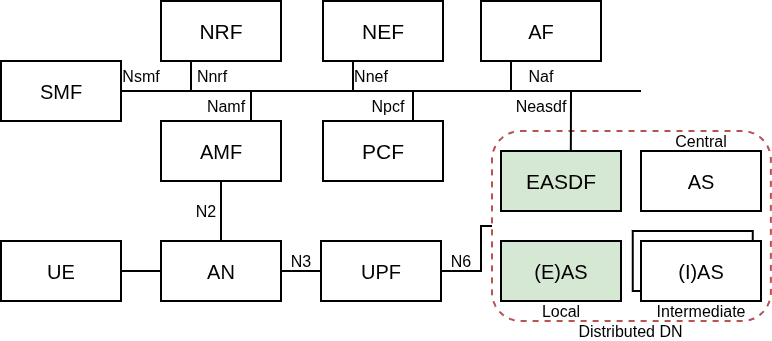}
    \caption{5G Edge Computing Architecture (adapted from~\cite{ts23548}).}
    \label{fig:easdf}
\end{figure}

\begin{figure}[t]
    \centering
    \includegraphics[width=0.9\columnwidth]{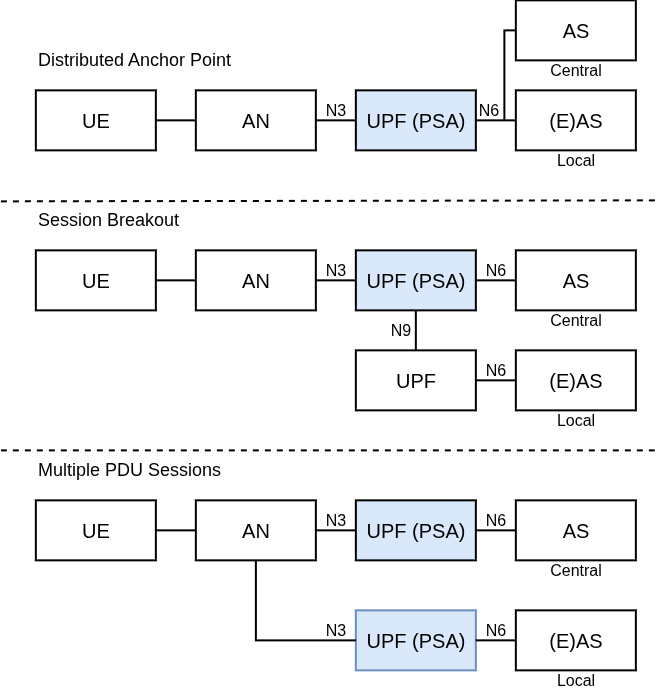}
    \caption{5G Edge Computing Connectivity Models (adapted from~\cite{ts23548}).}
    \label{fig:models}
\end{figure}

5G natively supports the edge computing paradigm through a set of features specified in~\cite{ts23501,ts23548}. Fig.~\ref{fig:easdf} shows the reference architecture. The DN can be distributed over multiple sites (e.g., a central and a local one), and is reachable from the User Plane Function (UPF) through the standard N6 interface. The local site can host an Edge Application Server (EAS), while the central site hosts the origin Application Server (AS). Intermediate Application Servers (IASs) may be employed to distribute the content to intermediate points between the local and central part of the DN.
A key enabler is the Edge Application Server Discovery Function (EASDF), introduced in 3GPP Release~17~\cite{ts23548,ts29556}. The EASDF acts as a DNS resolver for the UE and handles DNS messages according to the instructions received by the SMF: in this way, the traffic of a UE can be steered towards a local EAS, one or more intermediate servers, or the origin, based on criteria such as QoS, load balancing, or content availability\cite{carota2024easdf}.
Three connectivity models are defined in~\cite{ts23548}, as shown in Fig.~\ref{fig:models}:
\begin{itemize}
    \item \textbf{Distributed Anchor Point}: the PDU Session Anchor (PSA) UPF is located at the local site, so that all the traffic of the PDU session exits towards the DN through N6. Central ASs are reached from the local site.
    \item \textbf{Session Breakout}: the PSA is centralized, and an additional local UPF, acting as Uplink Classifier (ULCL) or Branching Point (BP), breaks out selected traffic towards the local EAS. The two UPFs are connected through N9.
    \item \textbf{Multiple PDU Sessions}: the UE establishes separate PDU sessions, anchored at different UPFs, for local and central services, e.g., based on UE Route Selection Policies (URSP).
\end{itemize}

In 5G-EMERGE, we adopted the distributed anchor point model, since it is transparent to both the UE and the 5GC, and thus does not require the availability of advanced features (e.g., ULCL insertion or URSP handling) in the core network or on the user side.
Regarding media playback, 3GPP considers DASH and HLS as the reference protocols for 5G Media Streaming (5GMS)~\cite{ts26501}. However, since the 5G system provides IP connectivity through PDU sessions and this is transparent to the application layer, any application-level streaming protocol is expected to work properly.

\subsection{5G Non-3GPP Access}\label{sec:n3gpp}
Since Release~15, the 5GC can also be accessed through non-3GPP access networks, such as Wi-Fi. These non-3GPP access networks can be classified as \textit{untrusted} or \textit{trusted} depending on whether the MNO trusts the access network or not~\cite{ts23501}. Each case relies on a dedicated Network Function (NF) acting as a gateway to the 5GC, i.e., the Non-3GPP Inter-Working Function (N3IWF) for untrusted access and the Trusted Non-3GPP Gateway Function (TNGF) for trusted access. However, in both cases, the communication relies on IPSec to establish a secure channel between the UE and the corresponding gateway function (i.e., N3IWF or TNGF). %
Since PDU sessions established over non-3GPP access are anchored at a UPF exactly as over 3GPP access~\cite{ts23548}, the edge computing concepts described above apply to both 3GPP and non-3GPP access.

\subsection{Multicast Delivery in 5G}\label{sec:mbs}

The 3GPP defines two main delivery methods for 5G Multicast and Broadcast Services (MBS)~\cite{ts23247}. In the \textit{individual} delivery method, the UPF receives a single copy of each packet from the source, replicates it, and forwards one copy to each UE over its already established PDU session. In the \textit{shared} delivery method, a single copy is forwarded to the NG-RAN, which delivers it to the UEs either in Point-to-Point (PTP) mode (i.e., with an individual copy per UE) or in Point-to-Multipoint (PTM) mode (i.e., with a single transmission over a shared radio channel). 
Since shared delivery relies on the NG-RAN, it is currently confined to 3GPP access, while only individual delivery over existing PDU sessions is also applicable to non-3GPP access. For this reason, the 5G-EMERGE ecosystem adopts the individual delivery method. %

\section{Testbed Description}\label{sec:testbed}
\begin{figure}[t]
    \centering
    \includegraphics[width=\columnwidth]{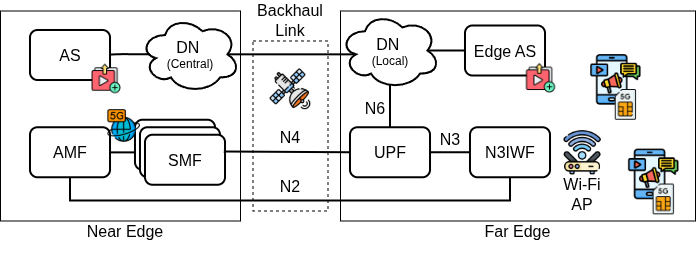}
    \caption{5G-EMERGE Distributed Testbed Architecture.}
    \label{fig:testbed}
\end{figure}

The testbed, shown in Fig.~\ref{fig:testbed}, is based on the distributed testbed described in~\cite{verde2026multicast}. The 5G system is split between a near edge, hosting the control plane functions of the HPE commercial 5GC, and a far edge, located at RomARS premises, hosting the UPF, the N3IWF, and the Wi-Fi Access Point (AP) through which UEs connect. The two sites are interconnected through (i)~a satellite backhaul based on the Starfish\footnote{https://www.mbigroup.it/en/advanced-telecommunications-systems/new-waveforms/} platform by MBI or (ii)~a terrestrial backhaul based on the Internet, depending on the test configuration. The DN is distributed as well: the central part hosts the origin AS, while a local part, at the far edge, is directly reachable from the UPF via N6 and hosts EASs (such as caches and replica servers) which can be fed by satellite multicast. The UPF acts as a PSA according to the distributed anchor point model, so that the PDU sessions of the UEs are anchored at the far edge.

UEs connect to the 5GC through untrusted non-3GPP access, via the Wi-Fi AP and the N3IWF. Both the N3IWF and the non-3GPP UE are proprietary implementations developed by the authors and deployed on separate interconnected Virtual Machines (VMs). Each VM has 2 vCPUs, 2GB RAM, 25GB SSD, and Ubuntu 22.04 with Linux Kernel v5.19. The UPF is a proprietary implementation as well, which applies the unicast Packet Detection Rules (PDRs), Forwarding Action Rules (FARs), and QoS Enforcement Rules (QERs) received from the HPE 5GC over N4~\cite{ts29244} for the corresponding PDU Session. To support multicast according to the 5G individual delivery method~\cite{ts23247}, the UPF includes an Internet Group Management Protocol (IGMP) Manager component, which (i)~verifies that the UE is authorized to access the multicast service, based on the rules received via N4, (ii)~handles the IGMP Join/Leave messages received from the UEs, and (iii)~dynamically installs/removes (at the kernel level) PDRs, FARs, and QERs required to replicate the multicast packets over the PDU session of each joined UE, as introduced in Section \ref{sec:mbs}. Since multicast data is carried over existing PDU sessions, the N3IWF acts as a relay, decapsulating packets received over the N3 interface and encapsulating them into the IPsec tunnels towards the UEs. %

In addition, the following services are deployed:
\begin{itemize}
    \item an origin server hosted on Amazon Web Services (AWS\footnote{https://aws.amazon.com/}), exposing the content via DASH/HLS. As a reference, the content is a live TV channel provided by RAI\footnote{https://www.rai.it/}, the Italian public broadcaster;
    \item a commercial CDN (Gcore\footnote{https://gcore.com/}), configured to fetch content from the AWS origin in case of a CDN miss;
    \item a Varnish Cache\footnote{https://www.varnish.org/} instance in the local part of the DN. In case of a cache miss, the content is fetched directly from the origin;
    \item a proprietary edge server (i.e., FTA Inverto\footnote{https://www.inverto.tv/}) in the local part of the DN, receiving the live content via DVB-MABR over the satellite backhaul link and streaming it in multicast UDP to the 5G users.
\end{itemize}

The video content is played in unicast through the Chromium\footnote{https://www.chromium.org/} web browser using the HNR reference player\footnote{https://testharness.media-streaming.testbed.5g-emerge.io/}, and the latency is computed from the HTTP Archive (HAR) files as the waiting time between each HTTP request and the first byte of the corresponding response (i.e., the Time-To-First-Byte, TTFB). On the other hand, multicast content is played with VLC media player\footnote{https://www.videolan.org/vlc/}, which automatically sends an IGMP Join Request when playback starts. The join latency is measured at the UE side as the time difference between the sending time of the join request and the arrival time of the first multicast packet. The throughput is measured over the N6 interface of the UPF with Wireshark.

\section{Results}\label{sec:results}

\subsection{Latency: CDN vs. Origin vs. 5G DN Cache}\label{sec:test1}
In the first test, one player is configured to fetch the DASH content directly from the AWS origin server. Three independent runs of about 180~s are performed, and the TTFB is measured as described in Section \ref{sec:testbed}. 

In the second test, the content is fetched from the Gcore CDN. Also in this case, three independent runs of about 180~s are executed. In each run, two 5G UEs play the same DASH content, with the second UE starting the playback after the first one. In this way, the requests of the first UE result in CDN misses, while those of the second UE likely result in CDN hits. In fact, this expected behaviour has been validated by checking the cache status headers of the HTTP responses. 

\begin{figure}[!t]
    \centering
    \includegraphics[width=\columnwidth]{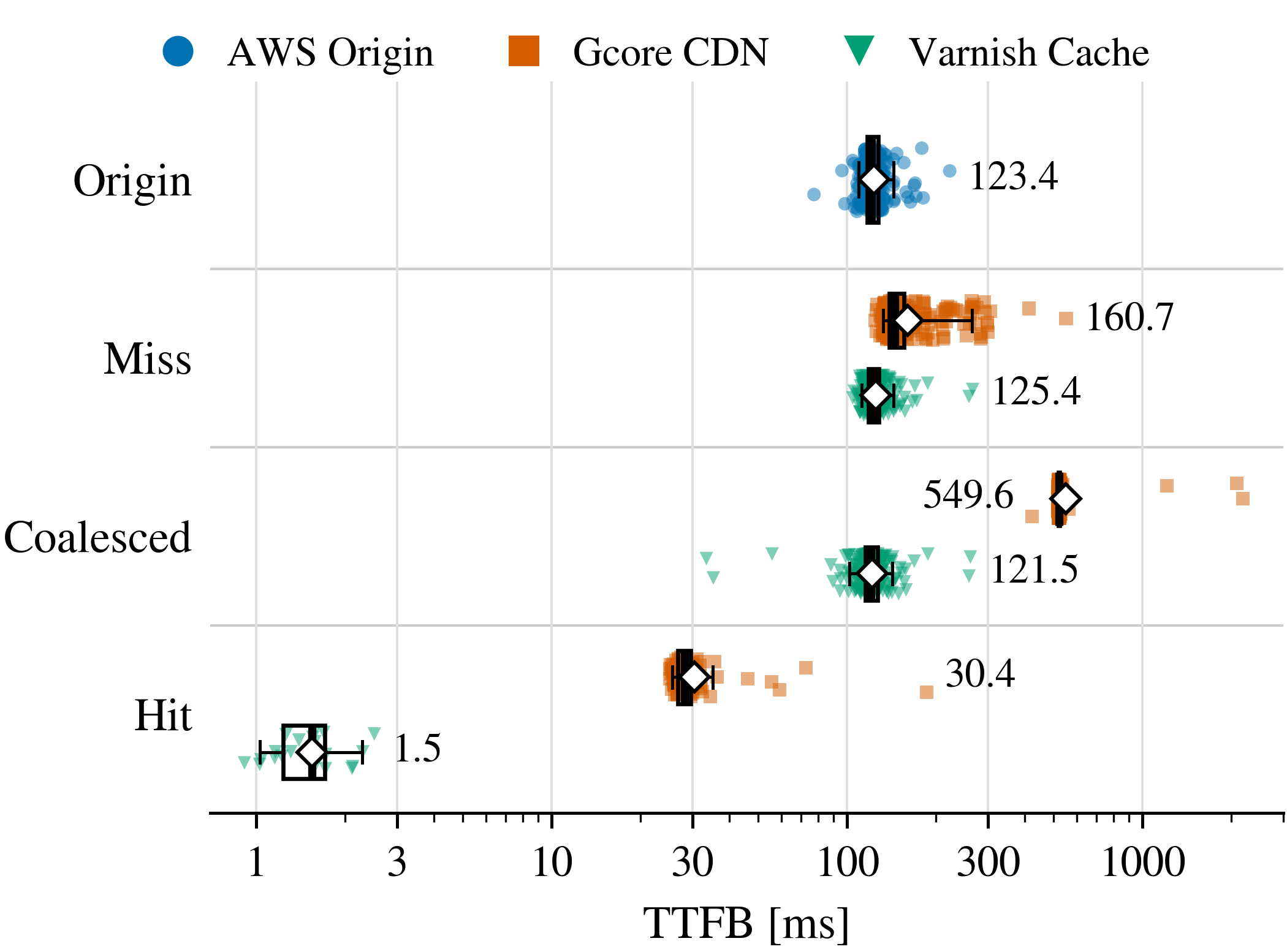}
    \caption{TTFB of DASH segments: AWS origin vs. Gcore CDN vs. Varnish cache in the 5G local DN: hit, miss, and coalesced requests.}
    \label{fig:results-ttfb}
\end{figure}

In the third test, the player fetches the content from the Varnish cache deployed in the 5G DN. A static DNS resolving rule is added to forward all HTTP requests to the cache, which in turn forwards each request to the origin in case of a miss. Two 5G UEs have been executed in order to play the content as in the second test case, thus expecting the same hit/miss behaviour. In all three tests, near and far edges are interconnected through a terrestrial backhaul, even though similar results have been experienced using satellite as backhaul, taking into account the additional delay introduced by the satellite network. Figure \ref{fig:results-ttfb} shows the aggregated results computed over all runs.
As a reference, the average Round-Trip Time (RTT) between the UE and the AWS origin, measured through 30 TCP SYN/SYN-ACK exchanges, is about 44.7~ms. Since the TTFB includes one network RTT and the server processing time, the remaining 78.7~ms (i.e., about 63.8\% of the average TTFB) is likely due to HTTP request processing at the origin.
In case of a CDN hit, the average latency is 30.4~ms, i.e., 75.4\% lower than fetching from the origin (123.4~ms). In case of a CDN miss, the latency (i.e., 160.7~ms) is 30.2\% higher than the origin one. This result is expected since the PoP must first fetch the segment from the origin before forwarding it to the requesting client.
This test also revealed an unexpected, implementation- or configuration-related behaviour. In more detail, when a UE has requested a segment that is not yet in cache, and another UE requests the same segment before the response is received, the CDN puts the second request on hold and polls the waiting queue every about 500~ms. Then, if the requested content is available in cache, it is served. In our case, this resulted in an average latency of 549.6~ms for the CDN coalesced requests. This mechanism, known as \textit{request coalescing}, prevents the origin from being overloaded when a popular object expires or is not yet in cache (i.e., the \textit{thundering herd} problem). However, it significantly increases the latency perceived by users.
In case of a cache hit, the average latency drops to 1.5~ms, i.e., lower not only than the origin but also than the CDN hit case, since the local DN cache is closer to the UE than the CDN PoP. Moreover, no request was put on hold as in the CDN hit case, since the Varnish cache handles concurrent requests for the same object in an event-driven fashion: simultaneous requests are queued and, as soon as the object becomes available, the waiting clients are notified and served. In fact, in case of cache coalesced requests, the average latency (i.e., 121.5~ms) is slightly lower than fetching the content directly from the origin (i.e., 123.4~ms). Finally, it is worth noting that, in case of a cache hit, the traffic on the backhaul link connecting the local and remote parts of the DN is also avoided. 

\subsection{Multicast Delivery from the 5G DN}\label{sec:test3}
\begin{figure}[!t]
  \centering
  \subfloat[Unicast (HTTP)\label{fig:thr_unicast}]{%
    \includegraphics[width=\columnwidth]{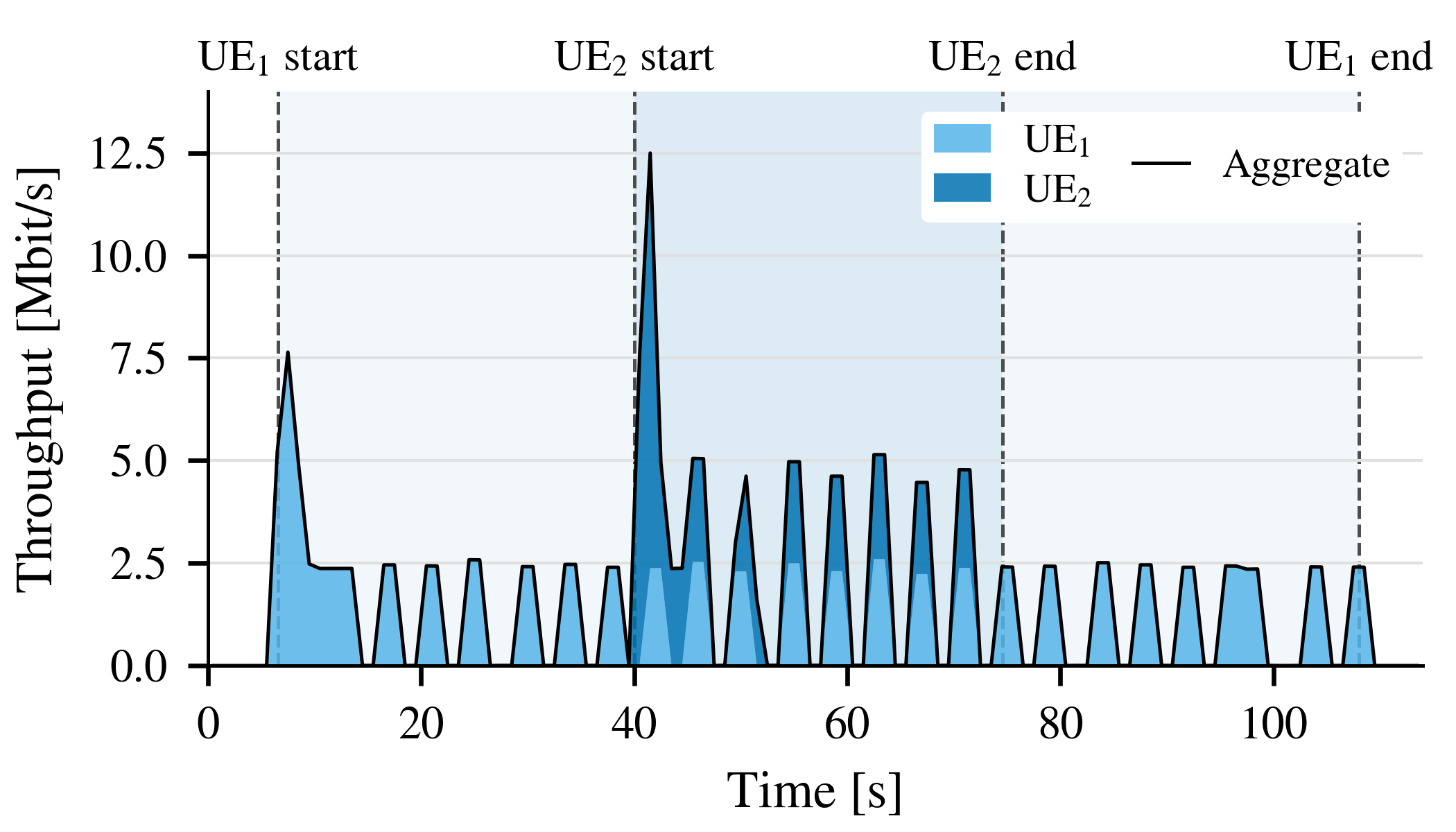}}\\
  \subfloat[Multicast (UDP)\label{fig:thr_multicast}]{%
    \includegraphics[width=\columnwidth]{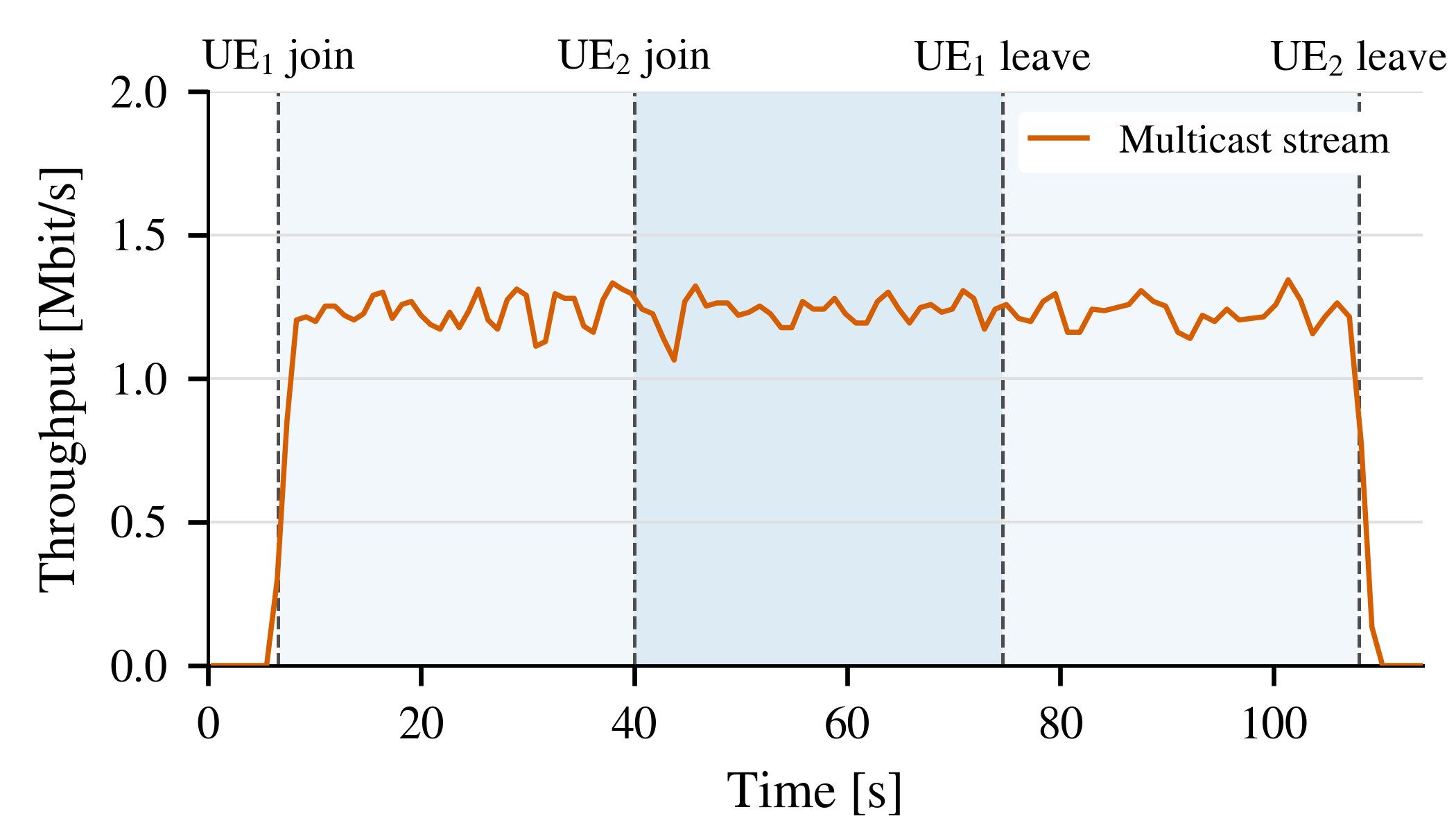}}
  \caption{Downlink traffic load over N6 interface.}
  \label{fig:throughput}
\end{figure}
In the fourth test, the two UEs play the same content from the FTA edge server located in the local part of the 5G DN using VLC. In this test, the edge server fetches the content from the origin via DVB-MABR over the satellite backhaul, and re-transmits it at the maximum available quality to the 5G UEs in (i) multicast UDP and (ii) unicast HTTP. Fig.~\ref{fig:throughput} shows the results in terms of throughput measured on the UPF N6 interface. Similar results are also observed with a terrestrial backhaul link.
As shown, when no UE has joined the multicast group or started video playback in unicast (i.e., the initial seconds in both Figs.~\ref{fig:thr_unicast} and~\ref{fig:thr_multicast}), the UPF receives no traffic on N6. When the first UE joins the multicast group (i.e., at $t=7$~s of Fig.~\ref{fig:thr_multicast}), an average throughput of about 1.2~Mbit/s is observed on N6. This rate remains stable and does not change when the second UE joins ($t=40$~s), nor when one of the two UEs leaves ($t=75$~s). This is the main difference with respect to the unicast case of Fig.~\ref{fig:thr_unicast}, where, as expected, the average throughput over N6 doubles (from about 1.2 to 2.4~Mbit/s) while the two UEs request the content from the edge server. Furthermore, Fig.~\ref{fig:thr_unicast} shows that unicast playback start causes a short start-up burst as the player fills its buffer. Finally, when the last UE leaves ($t=108$~s), there is no more traffic on N6 in both unicast and multicast cases.
Concerning latency, we observed an average join latency of about 11.4 ms. As a reference, the average RTT measured over 30 ICMP exchanges from the UE to the edge server is about 1.0 ms. Thus, the remaining time (i.e., 10.4~ms) is due to processing time, likely needed by the UPF to install the PDR/FAR/QER rules and route the multicast traffic to the UE.

In general, the 5G-EMERGE multicast delivery strategy is beneficial for two reasons. First, satellite multicast distributes the same content to many far edges at the same time, which is inherently more scalable than unicast OTT, since it reduces the bandwidth usage on the backhaul link. Second, if the edge server transmits each packet only once to the UPF in multicast, which is in charge of performing the replication over the PDU sessions of the joined UEs, this reduces not only the bandwidth usage of the backhaul but also the load on the edge server. Therefore, assuming $n$ joined UEs in a single far edge, the traffic load over the N6 interface is reduced by a factor $1/n$ (i.e., to a constant value) with respect to unicast delivery.

\section{Conclusion and Future Work}\label{sec:conclusion}

In this paper, we experimentally evaluated the content delivery strategies of the 5G-EMERGE ecosystem. Results confirm that, to minimize the latency perceived by users, content must be brought as close as possible to them. This can be done transparently by means of a CDN or by exploiting the local part of the 5G DN, which is the closest point to the users in a 5G system. Deploying a cache or replica server in the local part of the DN not only reduces the latency with respect to a CDN, but also reduces the traffic on the backhaul, thus mitigating congestion and the overload of the origin. Moreover, 5G multicast individual delivery reduces the traffic on the UPF N6 interface, since the edge server transmits each content only once, instead of once per request as in unicast, enhancing scalability with the number of UEs. Finally, we also showed that request coalescing, while protecting the origin, can significantly increase the latency, and should be carefully configured in CDNs serving live content.

As future work, we plan to evaluate failover scenarios with multiple, possibly hybrid terrestrial/satellite backhaul networks, analyzing the loss probability and its impact on QoE, and dynamically selecting the content between the local and remote parts of the DN, CDN, and origin.

\section*{Acknowledgment}
The work described in this article was supported by ESA contract 400013171/22/NL/AF 5G-EMERGE project. Responsibility for the content resides in the author or organisation that prepared it and can in no way be taken to reflect the official opinion of ESA.

\bibliographystyle{IEEEtran}
\bibliography{references}

\end{document}